\documentclass[conference]{IEEEtran}
\usepackage{amssymb,relsize,graphicx,amsmath,psfig,cite,float,setspace,braket}
\usepackage[left=0.680in,right=0.673in,top=0.860in,bottom=1.07in]{geometry}
\usepackage{amssymb,relsize,graphicx,amsmath,psfig,cite,float,subfigure,setspace}
\usepackage{xcolor}
\usepackage{tabularx}
\usepackage{xcolor}
\usepackage{tabularx}
\usepackage{flushend}
\usepackage{algorithm}
\usepackage{algpseudocode}
\usepackage{subcaption}
\usepackage{subfigure}
\usepackage{bm}
\usepackage{algorithm}
\usepackage{amsmath}
\algnewcommand\algorithmicinput{\textbf{Input:}}
\algnewcommand\algorithmicinit{\textbf{Init:}}
\algnewcommand\algorithmicoutput{\textbf{Output:}}
\algnewcommand\INPUT{\item[\algorithmicinput]}
\algnewcommand\Init{\item[\algorithmicinit]}
\algnewcommand\OUTPUT{\item[\algorithmicoutput]}
\newcommand{\MSE}{\mathrm{MSE}}
\newcommand{\tr}{\mathrm{tr}}

\newcommand{\Cov}[1]{\mathbf{C}_{\mathbf{#1}}}
\newcommand{\Exp}[1]{\mathbb{E}\!\left[#1\right]}

\newcommand{\bi}{\begin{itemize}}
\newcommand{\ei}{\end{itemize}}

\newcommand{\be}{\begin{enumerate}}
\newcommand{\ee}{\end{enumerate}}
\newcommand{\bd}{\begin{description}}
\newcommand{\ed}{\end{description}}
\newcommand{\bc}{\begin{center}}
\newcommand{\ec}{\end{center}}
\newcommand{\bt}{\begin{tabbing}}
\newcommand{\et}{\end{tabbing}}
\newcommand{\bfig}{\begin{figure}}
\newcommand{\efig}{\end{figure}}
\newcommand{\beq}{\begin{equation}}
\newcommand{\beqarr}{\begin{eqnarray}}
\newcommand{\beqarrn}{\begin{eqnarray*}}
\newcommand{\eeq}{\end{equation}}
\newcommand{\eeqarr}{\end{eqnarray}}
\newcommand{\eeqarrn}{\end{eqnarray*}}
\newcommand{\bflr}{\begin{flushright}\vspace{-0.2in}}
\newcommand{\eflr}{\end{flushright}}
\newcommand{\bsub}{\begin{subequations}}
\newcommand{\esub}{\end{subequations}}
\newcommand{\barr}{\begin{array}}
\newcommand{\earr}{\end{array}}
\newcommand{\nn}{\nonumber}

\def\undb#1{\mbox{\bf{#1}}}

\def\BibTeX{{\rm B\kern-.05em{\sc i\kern-.025em b}\kern-.08em
		T\kern-.1667em\lower.7ex\hbox{E}\kern-.125emX}}

\begin{document}

\title{\huge{Multi-User Symbol Detection with XL Reception:\\ Dynamic Metasurface Antennas with Low Resolution ADCs}\vspace{-0.3cm}}

\author{ \IEEEauthorblockN{Rahul K. Pal\IEEEauthorrefmark{1},  
Soumya P. Dash\IEEEauthorrefmark{1}, Barathram Ramkumar\IEEEauthorrefmark{1}, and George C. Alexandropoulos\IEEEauthorrefmark{2}}
\IEEEauthorblockA{\IEEEauthorrefmark{1}School of Electrical and Computer Sciences, Indian Institute of Technology Bhubaneswar, Khordha, Odisha, India \\
\IEEEauthorrefmark{2}Department of Informatics and Telecommunications, National and Kapodistrian University of Athens, 16122 Athens, Greece \\
{e-mails: \{s26ec09004, spdash, barathram\}@iitbbs.ac.in, alexandg@di.uoa.gr}}
\vspace{-0.6cm}
}

{}
\maketitle
\begin{abstract}
Dynamic Metasurface Antennas (DMAs) have been recently proposed as a cost- and energy-efficient front-end solution for eXtremely Large (XL) antenna array systems, supporting scalable Analog and Digital (A/D) beamforming while using a reduced number of Radio-Frequency (RF) chains. This array architecture is commonly realized as partially connected hybrid A/D beamformers, in which non-overlapping subarrays are linked to separate RF chains, each attached to a waveguide hosting multiple metamaterials. In this work, we study uplink multi-user communications where each RF chain of an XL DMA receiver is equipped with a $b$-bit resolution Analog-to-Digital Converter (ADC). We cast a Mean Squared Error (MSE) minimization problem for the design of the hybrid A/D combiner aimed at multi-user symbol detection, which is intrinsically non-convex due to the structural constraints imposed by the DMA hardware. By exploiting the Bussgang decomposition and a tractable modeling framework, we propose an efficient joint design of the hybrid A/D combining parameters. Our numerical evaluations showcase that XL DMA receivers can perform highly accurate multi-user symbol detection, revealing attractive trade-offs between hardware complexity and MSE performance.  
\end{abstract}
\section{Introduction}
In recent years, massive Multiple-Input Multiple-Output (mMIMO) has been widely recognized as one of the most impactful technologies for improving spectral efficiency and reliability through powerful spatial multiplexing and beamforming~\cite{10232975,allocated2014massive}. Despite these benefits, practical deployment is constrained by the reliance on a large number of Radio-Frequency (RF) chains \cite{mo2017hybrid}, 
resulting in high power consumption, cost, and complex transceiver architectures. These challenges are further exacerbated at millimeter-Wave (mmWave) and sub-Terahertz (sub-THz) bands due to the resulting  inter-antenna spacing and stringent form-factor requirements, limiting mMIMO scalability \cite{Faisal2020THZ}. 
Consequently, achieving a favorable tradeoff between performance, energy efficiency, and hardware complexity remains difficult. 

Dynamic metasurface antennas (DMAs) have recently emerged as a viable alternative antenna architecture, enabling analog beamforming, combining, and antenna selection with a substantially smaller number of RF chains \cite{shlezinger2020dynamic, Smith_2017, Chen2024, 10935678}.
DMAs actively reduce hardware complexity while preserving multiplexing/beamforming gains, offering a scalable and cost-effective solution \cite{Williams2022}.
Initial studies on DMAs presented in \cite{shlezinger2020dynamic,lan2020WSU}
highlighted their architecture, working principles, and showcased comparable performance with conventional mMIMO systems, while employing substantially fewer RF chains. Subsequent studies \cite{Kimaryo2024dbeamforming, 10938788} explored beamforming, achievable rate, and capacity in DMA-based uplink/downlink communication systems under flat and frequency-selective fading. Channel estimation techniques accounting for metasurface-induced frequency selectivity, waveguide propagation effects, and dimensionality reduction induced by the reduced number of RF chains were proposed in \cite{Rezvani2024cedma}. Energy-efficient designs for DMA-based XL MIMO systems were explored by jointly optimizing beamforming and power allocation to minimize power consumption and complexity in~\cite{9847609}. Furthermore, DMA-based MIMO orthogonal frequency-division multiplexing receivers with bit-constrained Analog-to-Digital Converters (ADCs) were studied in~\cite{Wang2021DMAMIMOOFDM}. A DMA-based receiver with $1$-bit ADC for multi-user uplink communications was proposed in~\cite{Gav2024}. Furthermore, DMA-equipped Base Stations (BSs) supporting near-field multi-stream transmissions were investigated in~\cite{Qiu2026LRDMA, Gavras2024}.

Most prior research on DMAs, including the above-mentioned ones, emphasize on sum-rate maximization, energy efficiency, and channel estimation. However, the important aspect of signal detection under practical hardware limitations at the receiver end has been largely overlooked. Motivated by this research gap, in this paper, we investigate uplink multi-user communications employing DMA-based reception with $b$-bit resolution ADCs. Using the Bussgang decomposition, we propose a framework based on Alternate Optimization (AO) for reliable multi-user symbol detection, by casting the detection task as a Mean Squared Error (MSE) minimization problem. The presented method simultaneously facilitates Analog and Digital (A/D) combining/processing, explicitly incorporating the structural properties of the DMA and the effects of quantization. Numerical results showcase enhanced detection performance across different ADC resolutions, and reveal inherent trade-offs among quantization levels, the number of RF chains, and the number of DMA elements.

{\em Notations:} Boldface letters, such as $\mathbf{A}$ and $\mathbf{a}$, denote matrices and vectors, respectively. $\mathbf{a}^T$ and $\mathbf{a}^\dagger$ output the transpose and the Hermitian (conjugate transpose) of $\mathbf{a}$, respectively. $\text{tr}\{\mathbf{A}\}$ gives the trace of $\mathbf{A}$ and $\lVert \cdot \rVert$ denotes the $\ell_2$-norm operator. A complex Gaussian distributed random vector with a mean vector $\bm{\mu}$ and covariance matrix $\mathbf{K}$ is denoted as $\mathcal{CN}\left( \bm{\mu}, \mathbf{K} \right)$. $\mathcal{R}(\cdot)$ and $\mathcal{I}(\cdot)$ represent the real-part and imaginary-part operators, respectively, $(\cdot)^*$ is the complex conjugate operator, and $\jmath\triangleq\sqrt{-1}$ is the imaginary unit. $\mathbb{E} [\cdot]$ is the statistical expectation operator, $\mathbf{0}_{N \times 1}$ and $\mathbf{1}_N$ are $N \times 1$ vectors of zeros and ones, respectively, $\mathbf{I}_N$ is the $N\times N$ identity matrix, $\mathrm{diag}(\mathbf{a})$ outputs a diagonal matrix with the elements of $\mathbf{a}$ as the diagonal elements, $\mathrm{blkdiag} \left[\mathbf{a}_1,\ldots,\mathbf{a}_N \right]$ denotes a block-diagonal matrix of the vectors $\mathbf{a}_1,\ldots, \mathbf{a}_N$, and $\otimes$ indicates the Kronecker product operator.
\section{System Model}
We consider the uplink of a multi-user communication system, where $K$ single antenna users are served by a multi-antenna BS equipped with a DMA comprising an array of $N \gg K$ response-tunable metamaterials. Each $N_e$-element distinct group of these metamaterials is placed on a waveguide in an one-dimensional arrangement, collectively termed as microstrip~\cite{shlezinger2020dynamic}. The total number of vertically arranged microstrips is $N_v$ microstrips, implying that $N = N_v N_e$. We denote the multi-user signal vector as $\mathbf{s} \triangleq \left[ s_1, \ldots, s_K \right]^T $ $\in {\mathbb{C}^{K \times 1}}$, where $s_k$ is the symbol transmitted by the $k$-th user ($k = 1, \ldots, K$), with an average transmit power represented by $P_s$.
The $N$-element signal received at the DMA can thus be mathematically expressed in baseband as:
\beq
 \mathbf{r} \triangleq \mathbf{H}\mathbf{s} + \mathbf{n},
 \label{eq:r}
\eeq
where $\mathbf{H} \triangleq \left[\mathbf{h}_1, \mathbf{h}_2, \ldots, \mathbf{h}_N \right]^T \in \mathbb{C}^{N \times K}$ denotes the multi-user channel matrix, with each element of \undb{H} being independent and identically distributed (i.i.d.) and following a zero-mean complex Gaussian distribution with unit variance. In addition, $\mathbf{n} \triangleq \left[ n_1, \ldots, n_N \right]^T \in \mathbb{C}^{N \times 1}$ denotes the noise vector distributed as $\undb{n} \sim {\mathcal{CN}} \left(\undb{0}_{N \times 1}, \sigma_n^2 \undb{I}_N \right)$. We define the average Signal-to-Noise Ratio (SNR) as $\Gamma_{\text{av}} \triangleq P_s/\sigma_v^2$. 

Each microstrip is attached to a reception RF chain, each including an ADC with $b$-bit resolution for both the in-phase and quadrature-phase components. The received analog signal $\mathbf{r}$ propagates through the waveguide, experiences attenuation, and phase shifts. We model this amplitude and phase distortion via $ \mathbf{A} \in \mathbb{C}^{ N \times N }$, a diagonal matrix with entries capturing the effect of signal propagation inside the waveguide as~\cite{Gavriilidis2025}:
\beq
\left[\undb{A} \right]_{\left(i-1\right)N_e + l, \left(i-1\right)N_e + l}
= \frac{e^{\left(\alpha + \jmath \beta \right) d_{i,l}}}{\sqrt{N_e}} \, , 
\label{eq:A}
\eeq
where $\alpha$ is the attenuation, $\beta$ represents the propagation constant, and $d_{i,l}$ represents the distance from the output port to the $l$-th element on the $i$-th microstrip ($i=1, \ldots, N_v$ and $l=1, \ldots, N_e$). In the considered uplink scenario, the signals are captured by the DMA elements and propagate along each microstrip to the RF chains, exhibiting distance-dependent attenuation and phase shift characterized by $d_{i,l}$. 

Each element of $\mathbf{r}$ in~\eqref{eq:r} is multiplied by a tunable analog weight $q_{i,l}$, which denotes the weight applied to the $l$-th element on the $i$-th microstrip of the DMA. The output signal at each microstrip is then formed by the superposition of the contributions from its constituent elements. The output of the $N_v$ microstrips is thus obtained as $\mathbf{y} \triangleq \mathbf{Q}\mathbf{A}\mathbf{r}$ $ \in \mathbb{C}^{N_v \times 1}$, where the analog combining matrix $\mathbf{Q} \triangleq \mathrm{blkdiag} \left[\mathbf{q}_1^{\dagger}, \ldots, \mathbf{q}_{N_v}^{\dagger} \right] \in \mathbb{C}^{N_v \times N}$ consists of the tunable weights of the DMA elements. Here, each $\mathbf{q}_i \triangleq \left[q_{i,1}, \ldots, q_{i,N_e} \right]$ comprises the coefficients of each element $q_{i,l}$ and is subject to the following constraints~\cite{10935678}:
\beq
q_{i,l} \in \mathcal{F} \triangleq
\left\{ 0.5\left(\jmath + {e}^{ \jmath \theta}\right)  
\middle | \theta \in \left[ 0,2\pi \right) \right\}.
\label{eq:q}
\eeq
The $N_v$ microstrip outputs are fed to respective RF chain, each including a pair of $b$-bit resolution ADCs that separately quantize the in-phase and quadrature components. 
The quantized output of each $i$-th RF chain is $z_i \triangleq \mathcal{Q}_b \left(\mathcal{R} \left( y_i \right) \right) + \jmath \mathcal{Q}_b \left( \mathcal{I} \left( y_i \right) \right)$, where $\mathcal{Q}_b \left(\cdot\right)$ is a $b$-bit uniform real value quantizer.
Since the DMA performs analog beamforming over a large number of antennas, using the central limit theorem, $\mathbf{y}$'s distribution can be approximated as Gaussian~\cite{Wang2021DMAMIMOOFDM}, following which, we use the Bussgang decomposition \cite{Demir2021BussgangDecomp} to linearly represent the quantized outputs of the $N_v$ microstrips. Hence, the quantized output vector $\mathbf{z}$ can be expressed as:
\beq
\mathbf{z} \triangleq \mathbf{F}_b \mathbf{y} + \mathbf{g},
\label{eq:z}
\eeq
where $\mathbf{F}_b \triangleq \Cov{yz}^\dagger \Cov{y}^{-1} \in \mathbb{C}^{N_v \times N_v}$ is the Bussgang gain matrix, $\Cov{y} \triangleq \Exp{\mathbf{y y}^{\dagger}} $, and $\Cov{yz} \triangleq \Exp{\mathbf{y z}^{\dagger}}$. In addition, $\mathbf{g} \in \mathbb{C}^{N_v \times 1}$ is the uncorrelated quantization distortion noise satisfying the conditions: $\Exp{\mathbf{gy}^{\dagger}} = 0$ and $\mathbf{C_g} \triangleq \Exp{\mathbf{gg}^{\dagger}} = \Cov{z} - \mathbf{F}_b \Cov{y} \mathbf{F}_b^{\dagger}$, where $\Cov{z} \triangleq \Exp{\mathbf{z z}^{\dagger}} = \mathbf{F}_b \Cov{y} \mathbf{F}_b^{\dagger} + \Cov{g}$. The matrix $\mathbf{F}_b$ depends on the quantizer's characteristics and the input signal variance. For a real-valued $b$-bit quantizer, the thresholds  are $-\infty=\tau_0 < \tau_1 < \cdots < \tau_{2^b}=\infty$ and the reconstruction levels are $\ell_0,\ell_1,\ldots,\ell_{2^b-1}$. The Bussgang gain can be obtained in closed form under the assumption of a Gaussian input to the quantizer. As previously mentioned, in the XL antenna regime, the quantizer input $\mathbf{y}$ can be modeled as a zero-mean Gaussian vector variable with covariance matrix $\Cov{y}$, i.e., $\undb{y} \sim \mathcal{CN} \left(\mathbf{0}_{N_v \times 1}, \Cov{y}\right)$. For the special case when $\Cov{y} = \left(K \Gamma_{\text{av}} + 1 \right) \mathbf{I}_{N_v}$, it holds $\mathbf{F}_b = \rho_b  \mathbf{I}_{N_v}$ where the Bussgang gain coefficient, $\rho_b$, can be expressed as~\cite{Jacob2017}:
\begin{align}
\rho_b & = \mathrm{diag} \left( \Cov{y} \right)^{-1/2}
\sum_{i=0}^{2^{b}-1}
\frac{\ell_i}{\sqrt{\pi}}
\Bigg( \exp \left( -\tau_i^2 \mathrm{diag} \left( \Cov{y} \right)^{-1} \right) \nn \\
& \qquad \qquad \qquad \qquad
- \exp \left( -\tau_{i+1}^2 \mathrm{diag} \left( \Cov{y} \right)^{-1} \right) \Bigg) .
\label{eq:Fb}
\end{align}
For the infinite resolution case, i.e., for $b =\infty$, we have $\rho_b =1$, and the Bussgang matrix simplifies to $\mathbf{F}_b = \mathbf{I}_{N_v}$.
\section{Signal Detection and Problem Formulation}
Upon receiving the signal, the DMA-based receiver aims to estimate the transmitted signals from the multiple users. As shown in \eqref{eq:z}, the transmitted symbol vector $\mathbf{s}$ is received at the BS and processed through the DMA-based analog beamformer, followed by $b$-bit resolution ADCs and digital linear processing. Denoting the digital beamformer employed at the BS by $\mathbf{W} \in \mathbb{C}^{N_v \times K}$, the soft estimate $\hat{\mathbf{s}}$ of the transmitted symbol $\mathbf{s}$ is mathematically expressed as follows:
\beq
\mathbf{\hat{s}} \triangleq \mathbf{W}^{\dagger} \mathbf{z}.
\label{eq:s_hat}
\eeq
Following this soft estimate, we wish to optimally configure the DMA weights $\mathbf{Q}$ and the digital linear filter $\mathbf{W}$ such that the estimation error between the transmitted and detected signals is minimized. To this end, we formulate the following MSE minimization problem for the signal detection process:
\begin{align}
\min_{\mathbf{W},\mathbf{Q}} 
& \quad \Exp{ \left \lVert \mathbf{s}
- \mathbf{W}^{\dagger} \mathbf{z} \right \rVert^2 } \nn \\
\text{s.t.} & \quad 
\mathbf{Q} = \mathrm{blkdiag} \left[\mathbf{q}_1^{{\dagger}}, \ldots, \mathbf{q}_{N_v}^{{\dagger}} \right] \, , \,
\mathbf{q}_i \in \mathcal{F}^{N_e \times 1} \, \forall i.
\label{eq:mseopt}
\end{align}
After performing some straightforward mathematical manipulations, (\ref{eq:mseopt}) can be rewritten as follows:
\begin{align}
\min_{\mathbf{W}, \mathbf{Q}} & \quad \tr \{\Cov{s} \}
- 2\mathcal{R} \left(\tr\{\mathbf{W}^{\dagger} \Cov{zs}\}\right)
+ \tr\{\mathbf{W}^{\dagger} \Cov{z}\mathbf{W}\} \nn \\
\text{s.t.} & \quad 
\mathbf{Q} = \mathrm{blkdiag} \left[\mathbf{q}_1^{{\dagger}}, \ldots, \mathbf{q}_{N_v}^{{\dagger}} \right] \, , \,
\mathbf{q}_i \in \mathcal{F}^{N_e \times 1} \, \forall i,
\label{eq:opt}
\end{align}
where $\Cov{s} \triangleq \Exp{\mathbf{s}\mathbf{s}^{\dagger}}$ and
$\Cov{zs} \triangleq \Exp{\mathbf{z}\mathbf{s}^{\dagger}}$. It is noted that this optimization problem is nonconvex due to the structural constraints of the DMA receiver architecture. In particular, each constituent microstrip behaves as an independent waveguide, since no physical connections (nor couplings) exist among microstrips. Furthermore, the elements located on the same microstrip combine their absorbed signals. These hardware constraints significantly constrain the feasible set of DMA element weights, making the joint design of $\mathbf{Q}$ and $\mathbf{W}$ rather complex. In the sequel, we adopt the AO approach to obtain the optimal digital and analog combiners to minimize the system's MSE for multi-user symbol detection.
\subsection{Optimization of the Digital Combiner $\mathbf{W}$}
For a given DMA combiner matrix $\mathbf{Q}$, the digital combiner design problem can be formulated, using (\ref{eq:opt}), as follows:
\beqarr
\min_{\mathbf{W}} \quad \tr \{\Cov{s} \} - 2\mathcal{R}\left[\tr\{\mathbf{W}^{\dagger} \Cov{zs}\}\right] + \tr\{\mathbf{W}^{\dagger}  \Cov{z}\mathbf{W}\}.
\label{eq:optdig_1}
\eeqarr
It can be easily verified that the constituent objective function is a convex quadratic function of $\mathbf{W}$. Hence, the optimal digital combiner, $\mathbf{W}^{\text{opt}}$, is obtained by a unique minimizer characterized by the first-order optimality condition, yielding:
\beq
\mathbf{W}^{\text{opt}} = \Cov{z}^{-1}\Cov{zs} \, ,
\label{eq:Wopt}
\eeq
which is equivalent to the linear minimum MSE estimator of $\mathbf{s}$ from $\mathbf{z}$. Here $\Cov{zs} =  \mathbf{F}_b \mathbf{Q} \mathbf{A} \mathbf{H} \Cov{s}$ and $\Cov{y} = \mathbf{Q} \mathbf{A} \left( \mathbf{H} \Cov{s} \mathbf{H}^{\dagger} + \sigma_n^2 \undb{I}_N \right) \mathbf{A}^{\dagger} \mathbf{Q}^{\dagger}$, which results in $\Cov{z} = \mathbf{F}_b \mathbf{Q} \mathbf{A} \left( \mathbf{H} \Cov{s} \mathbf{H}^{\dagger} + \sigma_n^2 \undb{I}_N \right) \mathbf{A}^{\dagger} \mathbf{Q}^{\dagger} \mathbf{F}_b^{\dagger} + \Cov{g}$. Substituting $\Cov{z}$ and $\Cov{zs}$ into (\ref{eq:Wopt}) yields an explicit form of the optimal digital combiner $\mathbf{W}^{\text{opt}}$ as a function of $\mathbf{Q}$ and the Bussgang parameters. Furthermore, the expression of the system's MSE is obtained from \eqref{eq:optdig_1} as follows:
\begin{align}
\! \! \! \! \MSE & = \tr \left\{ \Cov{s} \right\}
- \tr \left\{ \left( \mathbf{F}_b \mathbf{Q} \mathbf{A} \mathbf{H} \Cov{s} \right)^{\dagger} 
\Big( \mathbf{F}_b \mathbf{Q} \mathbf{A}
\left( \mathbf{H} \Cov{s} \mathbf{H}^{\dagger} \right. \right. \nn \\
& \qquad \left. \left. \left.
+ \sigma_n^2 \undb{I}_N \right) \mathbf{A}^{\dagger} \mathbf{Q}^{\dagger} \mathbf{F}_b^{\dagger} + \Cov{g} \right)^{-1} 
\mathbf{F}_b \mathbf{Q} \mathbf{A} \mathbf{H} \Cov{s} \right\} .
\label{eq:mseoptq_1}
\end{align}
\subsection{Optimization of the Analog Combiner $\mathbf{Q}$}
It can be observed that the MSE expression in ({\ref{eq:mseoptq_1}}) is a non-convex function of $\mathbf{Q}$ due to the structural constraints on this matrix and the inverse operation. By using matrix quadratic transformation, this MSE objective can be efficiently transformed yielding the following tractable problem formulation:
\begin{align}
\label{eq:mseoptq_2}
\max_{\mathbf{Q}} 
& \quad 2 \Gamma_{\text{av}} \mathcal{R} 
\left( \tr \left\{ \left( \mathbf{F}_b \mathbf{QAH} \right)^{\dagger} \mathbf{\Phi} \right\} \right) \nn \\
& \quad - \tr \left\{ \mathbf{\Phi}^{\dagger} \left(\mathbf{F}_b \mathbf{Q} \mathbf{\Upsilon} \mathbf{Q}^{\dagger} \mathbf{F}_b^{\dagger}
+ \frac{\Cov{g}}{\sigma_n^2} \right) \mathbf{\Phi} \right\} \nn \\
\text{s.t.} & \quad \mathbf{q}_i \in \mathcal{F}^{N \times 1} \,\forall i ,
\end{align}
where $\mathbf{\Phi} \triangleq \Gamma_{\text{av}} \left( \mathbf{F}_b \mathbf{Q} \mathbf{\Upsilon} \mathbf{Q}^{\dagger} \mathbf{F}_b^{\dagger} + \left( \Cov{g}/\sigma_n^2 \right) \right)^{-1} \mathbf{F}_b \mathbf{QAH}$ and $\mathbf{\Upsilon} \triangleq \Gamma_{\text{av}} \mathbf{A H H^{\dagger} A^{\dagger} + A A^{\dagger}}$. 
However, this optimization problem is still non-convex due to the inverse relation on $\mathbf{Q}$. Thus, to obtain a problem that is more amenable to optimization, we reformulate the objective in terms of a stacking vector $\mathbf{q}$ that contains all the non-zero entries of $\mathbf{Q}$. By employing the vectorization operator and exploiting the standard Kronecker product identities, the \eqref{eq:mseoptq_2}'s objective can be further expressed as a quadratic function of $\mathbf{q}$, resulting in the reformulation:
\begin{align}\label{eq:mseoptq_3}
\max_{\mathbf{q}} 
& \quad 2\mathcal{R} \left\{ \Gamma_{\text{av}}
\left( \sum_{k=1}^{K} \left( \Phi_k^{\mathrm{T}} \mathbf{F}_b \otimes \mathbf{h}_k^{\dagger} \mathbf{A}^{\dagger}\right) \mathbf{B}\mathbf{q}\right) \right\} \nn \\
& \quad - \mathbf{q}^{\dagger} \mathbf{B}^{\dagger} 
\left( \sum_{k=1}^{K} \left( \mathbf{F}_b \Phi_k^{*} \Phi_k^{\text{T}} \mathbf{F}_b \otimes \mathbf{\Upsilon} \right) \right)  \mathbf{B} \mathbf{q} \nn \\
\text{s.t.} & \quad \mathbf{q}_i \in \mathcal{F}^{N \times 1} \,\forall i,
\end{align}
where $\mathbf{B} \triangleq \mathrm{blkdiag} \left[ \mathbf{b}_1, \mathbf{b}_2, \ldots, \mathbf{b}_{N_v} \right]$ defines the matrix of non-zero element vectors of the DMA weight matrix $\mathbf{Q}^{\dagger}$, satisfying the condition $\mathrm{vec} \left(\mathbf{Q}^{\dagger} \right) = \mathbf{B}\mathbf{q}$. Furthermore, $\Phi_k$ represents the $k$-th column of $\mathbf{\Phi}$. Consequently, the objective function in (\ref{eq:mseoptq_3}) can be reformulated, by utilizing $\mathrm{vec} \left(\mathbf{Q}^{\dagger} \right)$, the cyclic shift property of the trace operator, and the vectorization properties, as follows \cite{Wang2021DMAMIMOOFDM}:
\begin{align}
\max_{\mathbf{q}} &
\quad 2\mathcal{R} \left( \boldsymbol{\xi}^{\dagger} \mathbf{q} \right)
- \mathbf{q}^{\dagger} \boldsymbol{\Psi} \mathbf{q} \nn \\
\text{s.t.} & \quad \mathbf{q}_i \in \mathcal{F}^{N \times 1} \, \forall i,
\label{eq:mseopt_q2}
\end{align}
where $\boldsymbol{\xi}$ and $\boldsymbol{\Psi}$ are defined as:
\begin{align}
\boldsymbol{\xi} & = \Gamma_{\text{av}}
\left( \sum_{k=1}^{K} \left( \Phi_k \mathbf{F}_b \otimes \mathbf{h}_k^{\dagger} \mathbf{A}^{\dagger} \right) \mathbf{B} \right)^{\dagger}, \nn \\
\boldsymbol{\Psi} & = \mathbf{B}^{\dagger}
\left( \sum_{k=1}^{K} \left( \mathbf{F}_b \Phi_k^{*} \Phi_k \mathbf{F}_b \otimes \mathbf{\Upsilon} \right) \right) \mathbf{B}.
\label{eq:xi_psi}
\end{align}

It can be seen from (\ref{eq:mseopt_q2}) that $\mathbf{F}_b$ and $\Cov{g}$ are nonlinear in $\mathbf{q}$, a fact that makes their analytical treatment intractable. To address this, we exploit the asymptotic properties of the multi-user channel matrix $\mathbf{H}$ in the large-$K$ regime, as in \cite{Gavriilidis2025}. As the number of users $K \rightarrow \infty$, we obtain $\frac{1}{K}\mathbf{H}\mathbf{H}^{\dagger} \rightarrow 1$, it holds for the ratio: $\frac{1}{K}\frac{[\mathbf{H}\mathbf{H}^{\dagger}]_{n,n}}{[\mathbf{H}\mathbf{H}^{\dagger}]_{i,j}} \rightarrow \infty$ $\forall i,j \in \left\{1,\ldots, N \right\}$. This results in $[\Cov{y}]_{n,n} = \frac{N_e}{2} \left(K \Gamma_{\text{av}} + 1 \right)$, $\boldsymbol{\xi} = \Gamma_{\text{av}} \rho_b\left( \sum_{k=1}^{K} \left(\Phi_k  \otimes \mathbf{h}_k^{\dagger} \mathbf{A}^{\dagger} \right) \mathbf{B} \right)^{\dagger}$, and $\boldsymbol{\Psi} = \rho_b^2 \mathbf{B}^{\dagger} \left( \sum_{k=1}^{K} \left( \Phi_k^{*} \Phi_k \otimes \mathbf{\Upsilon} \right) \right) \mathbf{B}$. Consequently, for large values of $K$ or in the low SNR regime, it becomes $\Cov{g} = \left(1-\rho_b^2\right)\frac{N_e \left( K \Gamma_{\text{av}} + 1 \right)}{2} \mathbf{I}_{N_v} $ and the Bussgang matrix $\mathbf{F}_b = \rho_b \mathbf{I}_{N_v}$, where $\rho_b$ can be expressed as follows:
\beq
\rho_b \! = \! \! 
\sum_{i=0}^{2^{b}-1}
\sqrt{\frac{2\ell_i^2}{\pi N_e\left(K\Gamma_{\text{av}} + 1\right)}}
\left( \! e^{-\frac{2\tau_i^2}{N_e \left( K \Gamma_{\text{av}} + 1\right)}}
\! -e^{-\frac{2\tau_{i+1}^2}{N_e\left(K \Gamma_{\text{av}} + 1\right)}} \right) \! .
\label{eq:rho_b1}
\eeq
Thus, the optimization problem in (\ref{eq:mseopt_q2}) can now be solved employing the Lorentzian-constrained weights in~\eqref{eq:q}, where $\mathbf{q}$ takes the form $\mathbf{q} = \frac{1}{2}\left( \mathbf{u} - \jmath  \mathbf{1}_N \right)$ using the definition $\mathbf{u} \triangleq \left[ e^{j\theta_1}, \ldots, e^{j\theta_N}\right]$, with $\theta_i \in \left[ 0, 2 \pi \right)$ $\forall i \in \left\{1,\ldots,N \right\}$. Substituting this into~\eqref{eq:mseopt_q2}, yields the reformulation:
\begin{align}
\label{eq:mseopt_q3}
\max_{\mathbf{u}}
& \quad \mathcal{R} \left( \left(\boldsymbol{\xi}
+ \jmath \boldsymbol{\Psi} \mathbf{1}_N \right)^{\dagger} \mathbf{u} \right)
-\frac{1}{2} \mathbf{u}^{\dagger} \boldsymbol{\Psi} \mathbf{u} \nn \\ \text{s.t.} & \quad \left \lVert \mathbf{u}_i^* \mathbf{u}_i \right \rVert^2 = 1 \, \forall i .
\end{align}
It is evident that this problem remains non-convex owing to the equality quadratic constraint $\left \lVert \mathbf{u}_i^*\mathbf{u}_i \right \rVert^2 =1$ $\forall i $. However, this can be tackled by reformulating (\ref{eq:mseopt_q3}) in terms of a positive semidefinite matrix, similar to \cite{Gav2024}, as follows:
\beq
\max_{\mathbf{U}} \,\, \tr\{\mathbf{UV}\} \quad \text{s.t.} \quad  \mathrm{diag}\left(\mathbf{U}\right) =  \mathbf{I}_{N+1},\, \mathbf{U} \succeq \mathbf{0},
\label{eq:mseopt_q4}
\eeq
where $\mathbf{U}$ and $\mathbf{V}$ are $\left(  N +1 \right) \,\times \left(  N +1 \right)$ matrices given as:
\beq
\mathbf{U} \triangleq 
\begin{bmatrix}
\mathbf{u} \\ 1
\end{bmatrix}
\begin{bmatrix}
\mathbf{u}^{\dagger} & 1
\end{bmatrix} , \ 
\mathbf{V} \triangleq \frac{1}{2}
\begin{bmatrix}
-\boldsymbol{\Psi} &
2 \boldsymbol{\xi} + \jmath \boldsymbol{\Psi} \mathbf{1}_N \\
\left(2\boldsymbol{\xi} + \jmath \boldsymbol{\Psi} \mathbf{1}_N \right)^{\dagger} & 0
\end{bmatrix} .
\label{eq:UV}
\eeq
\begin{algorithm}[t!]
\caption{Proposed Joint Design of $\mathbf{W}^{\text{opt}}$ and $\mathbf{Q}^{\text{opt}}$}
\begin{algorithmic}[1]
\INPUT $K, N_v, N_e$, $P_s$, $\Gamma_{\text{av}}$, $\mathbf{H}$, and ADC parameters $\left(b, \tau_i, \ell_i\right)$.
\Statex \textbf{Initialization:}
\State Construct $[\mathbf{A}]_{(i-1)N_e+l,(i-1)N_e+l}$ using (\ref{eq:A}).
\State Initialize $\mathbf{Q}$ with random weights $q_{i,l}\in\mathcal{F}$ using (\ref{eq:q}).
\State Compute $\rho_b$ using \eqref{eq:rho_b1}, and set $\mathbf{F}_b = \rho_b \mathbf{I}_{N_v}$ and $\Cov{g} = \left(1- \rho_b^2\right)\frac{N_e\left(K \Gamma_{\text{av}} +1\right)}{2}$.
\For {$\mathrm{iter} = 1, 2, \ldots, \mathrm{iter_{max}} $}
\State Compute $\Cov{y}, \Cov{z}$, and $\Cov{zs}$.
\State Update $\mathbf{W} = \Cov{z}^{-1}\Cov{zs}$ using (\ref{eq:Wopt}).
\State Compute $\boldsymbol{\xi}$ and $\boldsymbol{\Psi}$ from (\ref{eq:xi_psi}).
\State Construct matrices $\mathbf{U}$ and $\mathbf{V}$ using (\ref{eq:UV}).
\State Solve the optimization problem (\ref{eq:mseopt_q3}) to obtain $\mathbf{Q}$. 
\State Update DMA weight $\mathbf{Q} =\mathrm{diag} \left[\mathbf{q}_1, \ldots, \mathbf{q}_{N_v} \right]^{\dagger} $.
\EndFor
\OUTPUT Optimal combiner matrices $\mathbf{Q}^\text{opt}$ and $\mathbf{W}^\text{opt}$.
\end{algorithmic}
\label{alg:jointAO_short}
\end{algorithm}

All in all, the initial nonconvex optimization problem in~\eqref{eq:mseoptq_2} has been converted into a solvable convex semidefinite relaxation problem, where we have replaced the rank-one constraint with a positive semidefinite constraint to obtain the optimal analog combiner $\mathbf{Q}^{\text{opt}}$. The overall AO-based approach solving~\eqref{eq:mseopt} to obtain $\mathbf{W}^{\text{opt}}$ and $\mathbf{Q}^{\text{opt}}$ is summarized in Algorithm~\ref{alg:jointAO_short}.
\section{Numerical Results and Discussion}
In this section, we investigate the performance of the proposed XL uplink communication system, comprising a DMA-based BS including $b$-bit resolution ADCs at its reception RF chains. In particular, we have evaluated the MSE performance and investigated the impact of key system parameters. The resolution of the ADC was varied for $b \in \left\{1, 2, 3, \infty \right\}$, where the assignment $b=\infty$ corresponds to an ideal infinite resolution ADC, which serves as the upper-bound benchmark. As previously described, the Bussgang decomposition was adopted to model the quantization process, and the system's MSE using the large-$K$ approximation has been assessed for the performance evaluation.

\begin{figure*}[t!]
\centering
\subfigure[MSE vs. average SNR,  $\Gamma_{\text{av}}$.]{
\includegraphics[height=1.8in,width=2.2in]{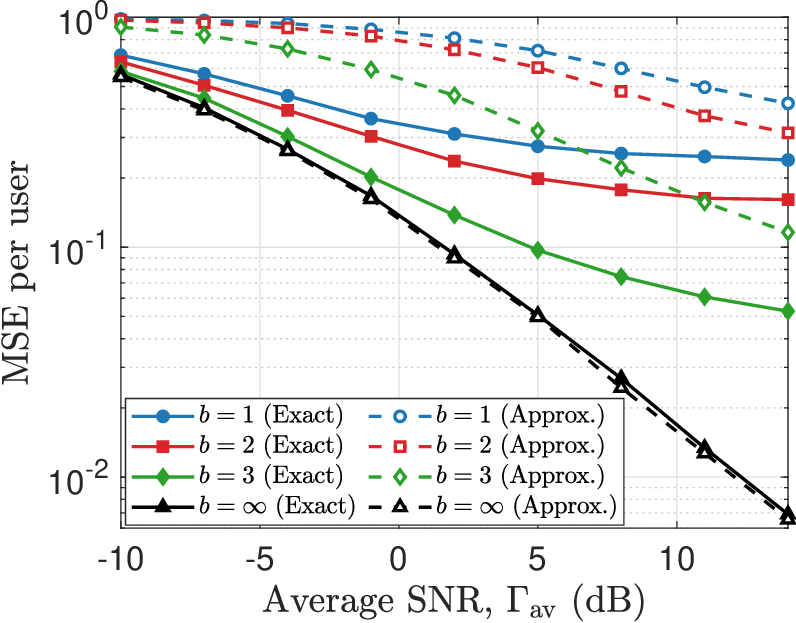}
\label{fig:1:a}
}\hfill
\subfigure[MSE vs. number of microstrips, $N_v$.]{
\includegraphics[height=1.8in,width=2.2in]{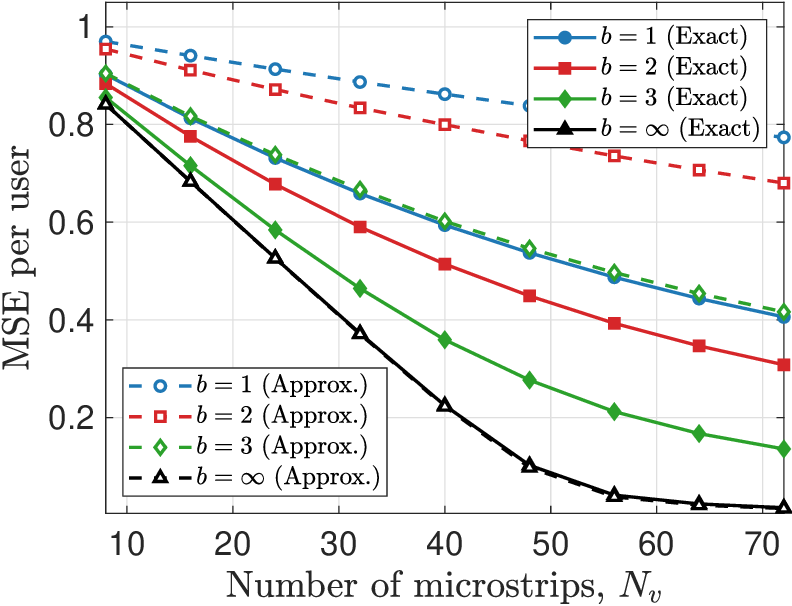}
\label{fig:1:b}
}\hfill
\subfigure[MSE vs. elements per microstrip $N_e$.]{
\includegraphics[height=1.8in,width=2.2in]{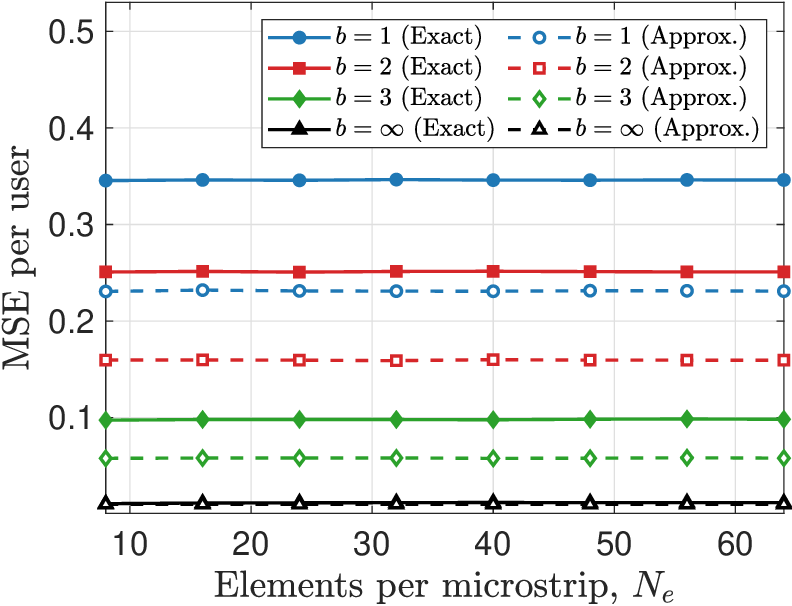}
\label{fig:1:c}
}
\caption{Comparison of the exact MSE and the MSE obtained using large-$K$ approximation versus (vs.): (a) $\Gamma_{\text{av}}$ for $N_e=200$, $N_v =10$, $K=40$; (b) $\Gamma_{\text{av}} = 5$ dB, $N_e=20$, $K=50$; and (c) $\Gamma_{\text{av}} = 5$ dB, $N_v=70$, $K=50$.}
\label{fig:MSE_DMA}
\vspace{-0.2cm}
\end{figure*}
Figure \ref{fig:1:a} presents the variation of the system's MSE with the average SNR, $\Gamma_{\text{av}}$, for $N_e = 200,\, N_v = 10$, and $K = 40$, under different ADC quantization levels. It can be observed that the various $b$-bit ADCs perform almost identically at low SNR values. This is attributed to the fact that the noise power exceeds the received signal covariance, and ADC quantization has a significant impact on the detection performance. As SNR increases, the effect of ADC resolution on the reduction in MSE becomes more pronounced, with the MSE decreasing at a faster rate as ADC resolution increases. In particular, the $1$-bit ADC exhibits the highest MSE due to the strong nonlinear distortion introduced by coarse quantization. Increasing $b$ reduces the distortion power and, therefore, improves the detection performance. Furthermore, the MSE for the large-$K$ approximation becomes tight with the exact MSE results with an increase in the ADC's resolution.

Figure \ref{fig:1:b} illustrates the variation of the MSE as a function of the number of microstrips $N_v$ for $\Gamma_{\text{av}} = 5$ dB, $ N_e = 20$, and $K = 50$. It is shown that increasing the number of microstrips significantly improves the system's performance. This improvement stems from the fact that each additional microstrip provides an extra spatial observation of the received signal, thereby enhancing the capability of the digital combiner to suppress multi-user interference and noise. When $N_v$ exceeds the number of users $K$, the receiver obtains sufficient spatial degrees of freedom to reliably separate the users, leading to saturation in the MSE.

Finally, Fig.~\ref{fig:1:c} depicts the MSE performance per user as a function of the number of microstrip elements $N_e$, for the setting $\Gamma_{\text{av}} = 5$ dB, $ N_v = 70$, and $K = 50$. The MSE remains invariant with increasing $N_e$ across all ADC resolutions, since additional elements per microstrip influence only the analog combining stage while leaving the dimension of the digital combiner input unchanged. This confirms that scaling the physical aperture of the XL DMA receiver neither degrades nor improves estimation performance. The resolution $b$ mainly sets the MSE level; as $b$ increases, the gap between exact and approximate values rapidly closes ($b \ge 3$).
\section{Conclusion}
In this paper, we investigated multi-user symbol detection with DMA-based reception utilizing low-resolution ADCs. The MSE criterion was used to formulate a hybrid A/D combiner design optimization problem that explicitly accounts for the effects of DMA-structural constraints and $b$-bit ADC resolution. Bussgang decomposition and a large-user approximation were used to tackle the difficult analog combiner design. This resulted in a tractable quadratic formulation, which was then solved by the semidefinite relaxation technique. Our numerical results demonstrated that, even with coarse quantization, the suggested framework enables accurate multi-user detection. Further, increasing the ADC resolution improves detection performance, with a $3$-bit ADC achieving almost perfect performance. Additionally, the number of microstrips has a significant impact on MSE, whereas the number of elements per microstrip results in a constant MSE, confirming that scaling up the antenna aperture does not degrade performance, and that near-optimal performance is achieved even with moderate numbers of microstrip elements.
\bibliographystyle{IEEEtran}
\bibliography{IEEEabrv,bibliography}
\end{document}